\documentclass[twocolumn,10 pt,showpacs,preprintnumbers,amsmath,amssymb]{revtex4}
\usepackage{amssymb}
\usepackage{graphicx}

\begin{document}

\title{Direct correlation between strengthening mechanisms and electrical noise in strained copper wires.}
\author{Natalia Bellido}
\author{Alain Pautrat}
\email{alain.pautrat@ensicaen.fr}
\author{Clement Keller}
\author{Eric Hug}
\affiliation{Laboratoire CRISMAT, UMR 6508 du CNRS, ENSICAEN et Universit\'{e} de Caen, 6 Bd Mar\'{e}chal Juin, 14050 Caen, France.}

\begin{abstract}
We have measured the resistance noise of copper metallic wires during a tensile stress.
The time variation of the main resistance is continuous
 up to the wire breakdown, but its fluctuations reveal the intermittent and heterogeneous character of plastic flow.
 We show in particular direct correlations
 between strengthening mechanisms and noise spectra characteristics.  
\end{abstract}

\maketitle

\section{Introduction}

 Crystalline metals are subject to dislocation nucleation and motion when they are submitted to stress. This leads to strain hardening,
 and ultimately to the breakdown of the material. This mechanical aspect has been driven into a more fundamental one 
when it has been realized that dislocation kinetics exhibit avalanche-like, scale free, properties \cite{Miguel},
 opening a route to a generic,
 to some extent material independent, physical approach of the process \cite{Zaiser}.
 At a macroscopic scale, however, a typical stress-strain curve appears still
 continuous due to the statistical averaging effect.
To reveal the motion of dislocations, microscopic samples can be used, and they display some characteristics of crackling noise,
 for example plastic strain bursts \cite{Dimiduk}.
Acoustic emission is also a dedicated probe of dislocation kinetics, sensitive to the microscopic strain mechanism.
 The statistical analysis of the data allows to build a probability density
 function which displays power-law distribution, as expected for avalanches.
 Recently, small scale self organized structures of acoustic events were observed \cite{Lebyo09}. 
 Concerning the electronic transport properties, it is known that the main resistivity can be influenced by the amount
 of dislocations which are present in the material \cite{dislo}.
 When thermally activated, dislocations are moving and lead to a resistance noise significantly larger than the thermal Johnson noise.
It was proposed that $1/f^{\alpha}$ with $\alpha \approx 2$ observed in metallic films is a characteristic of this dislocation noise,
 and is explained by a Poisson distribution of elementary voltage steps \cite{Dislocationnoise}. This
 is in contrast with the $1/f^{\alpha}$
with $\alpha \approx 1$ that generally arises in metal due to the quasi-equilibrium motion of defects
 (for example from the Dutta-Dimon-Horn process \cite{DDH}).
Bertotti et al also proposed that $1/f^{\alpha}$ noise with $\alpha \approx 4$ could appear for a frequency $f>f_0$ 
when the motion of dislocations is correlated \cite{Currentnoise}. In this model, $f_0$
is linked to the average duration of pulses due to dislocation motion. To generate plastic deformation responsible for the dislocation noise,
 the thermal induced strain generated
 by the lattice mismatch between a metallic film and the substrate can be used. With this technique, $1/f^2$ noise was observed but 
without $1/f^4$ contribution. This latter was supposed to be below the apparatus sensitivity \cite{Bert}.
 Low-frequency electrical noise, but close to $1/f$, was measured
 in carbon fibers subjected to tensile stress \cite{Pat},
 and in metal films submitted to different stress with a choice of different substrates \cite{Gior}. This $1/f$ noise spectrum
 can be attributed
 to a large distribution of activation energy in the range over $k_BT$ of elementary $1/f^2$ processes \cite{DDH}.
 If one measures the electrical noise during a tensile stress test,
 one can expect to obtain data directly correlated with the nucleation and motion of dislocations in the metal.
 Electrical noise was studied in stainless steel wires under
tensile stress, but the major contribution was a
narrow band noise which was interpreted as a manifestation of fluctuations in the piezoresistivity \cite{Haw}.
To our knowledge, only few works, devoted to the correlation between electrical properties and dislocation behaviour, take into account 
the structural heterogeneity of the dislocation patterns which takes progressively place inside the material during the 
strengthening (see for instance \cite{Narutani}).  
 \newline
In this paper, we report the measurements of voltage noise in a wire of pure copper during a monotonic tensile test.
 These measurements show that the changes between different stages of strain hardening and the resistance noise are directly correlated.
In particular, we report the observation of the $f^{-2}-f^{-4}$ cross-over typical of dislocation clustering,
 and some subtle effects which are hardly accessible by other techniques.  
 
\section{Experimental}

Experiments were performed on a polycrystalline copper wire of radius r=150 $\mu$m and purity of 99.99\% in weight. The main electrical
 resistivity was measured using a four probes method with an Adret A-103 current supply and a Keithley 199 Voltmeter. 
The measured value at $T=$ 300 $K$ was $\rho$ = 1.78 $\mu\Omega$.cm.
Regarding the noise set-up, the measured voltage was amplified by ultra low noise preamplifier (NF Electronics Instruments,
 model $SA-400 F3$, with an equivalent noise  $\approx$ 0.5 nV/Hz$^{1/2}$) enclosed in a thick screening box.
 Details of the set-up can be found in \cite{jo}. 
Signals were recorded with a $PCI-4551$ (National Instruments) analog input channel, anti-aliased
 and Fourier transformed in real time.
The quantity of interest is here the autopower spectra $S_{vv}(f)$ ($V^2/Hz$) of the voltage noise
 measured under a constant and noise free applied transport current of I=100 $mA$. 
Monotonic tensile tests were carried out using an Instron 5569 tensile test machine with suitable grips for thin wires. Grips were covered
 with insulating tape in order to isolate the wire and the test machine.
 The top grip was displaced at a constant rate of 10 $mm/min$ and the force was measured by an Instron static load cell of 50 $KN$.
 The total tested length was 250 $mm$,
 and voltage pads were separated by 187 $mm$.
 A special care was taken to ensure that the voltage spectrum is free from spurious contribution.
 In particular, a narrow band noise centered at a frequency of 25 Hz was observed,
 identified as coming from the vibration of the set-up and finally subtracted from the final spectrum. 
 The background electronic noise is dominated by the preamplifier noise ($\approx$ 0.5 nV/Hz$^{1/2}$) which is higher than
 the Johnson noise of the Copper wire at room temperature
 ($V_n=(4 K_B T R)^{1/2} \approx  0.03 nV/Hz^{1/2}$ at T=300 K).
 A schematic view of the set-up is shown in the figure 1. We have made several measurements. Some of them were unsuccessful because the contact noise
 was dominant (Ohmic contacts were degraded during the tensile test). Two of them were free from such contact noise, and gave similar results
that we report here.  

\section{Hardening behaviour of thin copper wires}

Typical rational stress ($\sigma$) - strain ($\epsilon$) curve of Cu is shown in the inset of Fig.2. The yielding stress stands around 180 MPa.
 The material exhibits for higher stresses
 three stages of strengthening typical of f.c.c metals behaviour \cite{Mecking 1977}.
 These stages can be conveniently studied when plotting the curve $\theta$ = $f(\sigma)$, with $\theta$ = d$\sigma$/d$\epsilon$ (Top of Fig.2).
 Stage I corresponds to microplasticity and appears for $\sigma$ values close to the yield stress.
 During this stage, plasticity progressively develops from grain to grain until a homogeneous state 
is reached. During this first stage the well favoured slip systems, $\left\{111\right\}<110>$ are activated in each grain. 
 The stage II is characterized by a constant value of $\theta$ as function of $\sigma$ \cite{Mecking 1977}.
 It is generally associated with
 the activation of less well oriented systems and a few cross slips \cite{Flinn et al. 2001}.
 Multiple gliding systems activation induces therefore the formation of heterogeneous dislocation
 structures like tangles, walls or cells. The stage III begins with the decrease of $\theta$ = $f(\sigma)$.
 It can be associated with a generalization of the cross-slip implying the annihilation of dislocations during
 dynamic recovery \cite{Essmann and Mughrabi 1979}.
TEM observations performed on f.c.c. metals, plastically strained at various strain levels, show that dislocation
 tangles develop at an early stage in the deformation \cite{Sumino et al. 1963}. Moreover, a well defined cell structure
 is always observed when the hardening rate reaches the second stage (see for instance \cite{Huang 1998} for Cu or recently
 \cite{Keller et al. 2010} for Ni). The dislocation cell size
 tends to decrease with an increase in strain, with a more pronounced way at the beginning of the second stage hardening.
 The mean free path of mobile dislocations is therefore determined by the cell size.

The total dislocation density $\rho^{d}$ monotonically grows whatever their structure from stages I to III. The stress $\sigma$
 is classically dependent of $\rho^{d}$ following the relationship \cite{Mecking and Kocks 1981}:
\begin{equation}
	 \sigma =\alpha \mu b M \sqrt{\rho^{d}}
\end{equation}
$M$ is the Taylor factor, $b$ the Burger's vector, $\mu$ the shear modulus and $\alpha$ a parameter taking
account the dislocation arrangement and the interactions between slip systems.

However, with the formation of cells, the local dislocation density may be significantly larger
 ($\rho^{d}_w$ inside the cell walls) or smaller ($\rho^{d}_c$
 in the cell interior) than $\rho^{d}$. The latter is therefore a consequence of a mixture law,
 following a composite model previously proposed \cite{Mughrabi 1987}:
\begin{equation}
	 \rho^{d} = f_w \rho^{d}_w + f_c \rho^{d}_c
\end{equation}

$f_w$ and $f_c$ are the area fractions of wall and cell interior, respectively, related through the relationship: $f_w + f_c = 1$.
 In particular, mobile dislocations are encountered inside the cells and their density rises more slowly than stored dislocations inside the cell walls.

Copper wires experienced a strong necking before the final fracture, which begins for true stresses higher than 300 $MPa$. We observed by scanning
 electron microscopy a section reduction of about 2/3 at the fracture point.

\section{Resistance noise due to dislocation kinetics}

We recall here the important points in the treatment of the dislocation noise, which was discussed for example in \cite {Bert}.
The main idea is that abrupt changes in the dislocation density generate resistance pulses which can be described as Poisson processes,
 similarly to a shot noise mechanism. 
 In the simplest case, it is assumed that the time dependent resistance is constituted by random and statistically independent (rectangular) pulses of
 height $\Delta R$ and duration $\tau$.
Some assumptions are required. $\tau$ is distributed according to an an exponential law $P(\tau)=\tau_0^{-1}exp(-\tau/\tau_0)$ and the resistance change is proportional to the average number of dislocations (per unit time) $N$ in the sample.
If the sample is biased with a constant current $I_0$,
the voltage spectrum takes finally the form:

\begin{equation}
S^{(1)}_{vv}= N (\Delta R)^2 I_0^2 / \pi^2 (1/\tau_0^2+4\pi^2 f^2)
\end{equation}

This is a typical Lorentzian spectrum for Poisson distributed elementary events \cite{vdz}.
The noise power is then proportional to $f^{-2}$ for frequencies higher than $f_0=1/(2\pi \tau_0)$.
 
If clustering takes place, the resistance pulses are correlated, and the power spectrum changes accordingly.
 Note that the model is in several aspects similar to some models of the magnetic Barkhausen noise \cite{mazetti}.
Considering bundles of dislocations, an averaged number of events per cluster $p$ can be defined,
 with $\tau$ the averaged lifetime between such events in the cluster. This correlation between
 the pulses introduces a new term in the power spectrum, and an approximate expression for the noise spectrum can be worked out \cite{Bert}:

\begin{equation}
S^{(2)}_{vv}=S^{(1)}_{vv}(1+2p(1-N \tau_0)^2)/(1+\omega^2\tau^2p^2(1-N\tau)^2)
\end{equation}
 
It can be simplified for an average number of pulses per clusters $p \gg 1$

\begin{equation}
S^{(2)}_{vv}=S^{(1)}_{vv}(1+2p)/(1+\omega^2\tau^2p^2)
\end{equation}

According to equation (3) and (5), the power spectrum can be rewritten in the compact form:
 
\begin{equation}
S^{(2)}_{vv}= N (\Delta R)^2 I_0^2 /(\pi^2 (f_0^{2}+f^2)) \times (1+2p/(1+(f/f^*)^2))
\end{equation} 

with $f^*=1/(2\pi\tau p$) the cut-off frequency of the clustered noise. Since $S^{(1)}_{vv} \propto f^{-2}$
 above the cut-off frequency $f_0$, $S^{(2)}_{vv}$ tends to 
a $f^{-4}$ behavior for large frequencies $f>f_0$. The characteristic
 frequencies follow the hierarchy $f^*<f_0$, i.e. the time constant of a cluster of dislocations ($\tau p$)
 is larger than the time constant of statistically independent dislocations.

\section{Results and Discussion}

The voltage noise $S_{vv}$ at a low frequency of 10 $Hz$ is shown as function of the strain in the bottom of the fig.2. 
Let us consider first the changes in the noise magnitude. A good correspondence between the variation of the
 voltage noise and the three stages of strengthening can be observed. 
As shown in the equation (3), the noise magnitude is proportional to $N (\Delta R)^2$, \textit{i.e.} to the rate of change of dislocation density
 in the sample and to the resistivity contribution of unit dislocation density \cite{Bert}.
 For low plastic stress levels (stage I), dislocation population mainly consists of isolated dislocations which slip along one crystallographic
 gliding system. During the stage II multiple slip systems are activated leading to an increase in dislocation density and to the formation of cell walls.
 This change in the density of dislocations may be
 at the origin of the noise magnitude increase during the stage II. When stage III is reached, the increase in $S_{vv}$ is reduced,
 this phenomenon can be explained by the occurrence of dislocation annihilation process, such as cross slips,
 which reduces the rate of dislocation accumulation \cite{Essmann and Mughrabi 1979}.
Close to the necking, around 310 $MPa$, the noise magnitude at 10 $Hz$ shows a bump.
 It is associated with a change of spectral shape which will be discussed later. 
 Finally, just before the fracture of the wire, the noise increases strongly. 
This occurs for stresses higher than about 320 $MPa$. This last noise is essentially white and intermittent (not shown here).
We propose that this increase in white noise is a resistance effect,
 revealed after the necking and explained by the fact that local section of the wire decreases and that small cracks start to develop
 just before the fracture. 

Let us consider the changes in the spectral shape during the tensile test. Some typical measured spectra are represented in figure 3.
 For low to moderate strains, the noise has a 
dominant $1/f$ character and a low magnitude (fig.3b).
 This behaviour changes when $\sigma \geq $ 200 $MPa$, close to the beginning of the stage II, with a progressive increase of the noise.
 At this stage, the noise spectrum changes to a $1/f^2$ form (fig.3c). The noise value tends to saturate for $\sigma \approx$ 225 $MPa$,
 apparently corresponding to the transition between the stage II and the stage III. Finally, close to the necking ($\sigma \approx 310 MPa$),
 the noise amount increases again. This time, both $f^{-4}$ and $f^{-2}$ components, separated by the characteristic frequency $f_{0}$,
 can be observed (fig.3d).
 
Regarding these spectral shapes, since stage II is defined by the formation of heterogeneous dislocation 
structures, two processes are expected: a slow one related to the clusters and a fast one related to dislocations 
inside clusters, so that according to Bertotti et al. \cite{Bert} both $f^{-2}$ and $f^{-4}$ behaviour should appear
 with the characteristic frequency of cross-over $f_{0}$. This frequency corresponds to an average duration $\tau_0\approx 2\pi/f_0$.
 This duration is that of the independent pulses due to dislocation motion. At the beginning, $f_{0}$ is not observed in the experimental
 window and only $f^{-2}$ behaviour is seen. But at the end of stage III, $f_{0}$  decreases and reaches the experimental window, 
so that both $f^{-2}$ and $f^{-4}$ are observed. The evolution of $f_{0}$ as function
 of stress is shown in Fig.4. A clear maximum can be observed around  311 $MPa$, \textit{i.e.} in the regime where the necking is thought
 to occur following the strain-stress behavior (Fig.2).
 Remembering that the necking is characterized by concentration of dislocations which suddenly localize in particular places,
 we think that the maximum of the frequency $f_0$ can be used as a new and precise probe of the necking of the wire. 
We show here that a high level of stress is necessary to reveal the $f^{-2}-f^{-4}$ cross over, likely associated to a large concentration of dislocations.
 This fact can explain why it was not observed in ref \cite{Bert}, where the stress was achieved by changing the temperature of
 an aluminium thin film clamped over a silicon substrate. In addition, in our case the study of dislocation noise is performed isothermally, 
avoiding the fact that dislocation dynamics under stress is temperature dependent, and allowing then a more direct analysis.
 
 An interesting aspect of plastic deformation is that the rearrangement of correlated dislocations shows similarities with an avalanche process \cite{Zaiser, Miguel}.
 Scaling laws typical of crackling noise can be then expected,
 and are actually observed by the acoustic emission experiments \cite{Weiss97}.
After plotting
 the number of the noisy events as function of the noise values in a log-log scale,
 we observe a non Gaussian probability distribution (Fig.5).
 More precisely,
  a power-law dependence $P(\delta V) \propto \delta V^{-\alpha}$ reasonably 
fits our data with $\alpha=1.5 \pm 0.1$.
 This exponent is close
 to the one measured when analysing the intermittent behaviour of the plastic response with acoustic emission,
 a result which indicates that dislocations move likely in a scale free fashion \cite{Miguel}.
The data scale whatever the strengthening stage (in the stage II and III),
 as observed for the acoustic emission experiments performed to study plastic activity \cite{Weiss97}.
Note however that we have limited statistics with our data, and a power-law behavior can not be rigorously proved.
 It remains that this result gives a good indication
 that our noise measurements
 and the acoustic emission experiments are observing similar statistical events.

\section{Conclusions}

In summary, we have observed resistance noise 
due to dislocation motion in a metallic wire submitted to stress.
 It was shown that different stages of the strengthening can be related to a specific
 behaviour of the voltage noise arising from the heterogeneity in the dislocation motion.
 Signatures of dislocation clustering are revealed.
 In addition to the technical interest of providing a non destructive characterization of strengthening stages and of the kinetics of plastic deformation,
this technique is an interesting tool to reveal hidden processes 
and to study avalanche-like dislocation kinetics, and is certainly complementary to acoustic emission experiments.

N.B and A.P thank Julien Aubril for his precious help during the data analysis.

\newpage
\begin{figure}[t!]
\begin{center}
\includegraphics*[width=8.0cm]{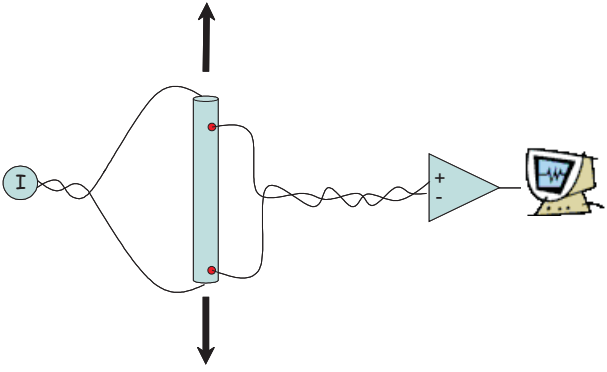}
\end{center}
\caption{(Color online) Schematic experimental set-up showing the copper wire with the noise free current supply,
 the contact pads, the preamplifier and the computer for the real time data acquisition.}
\label{fig.1}
\end{figure}

\begin{figure}[t!]
\begin{center}
\includegraphics*[width=8.0cm]{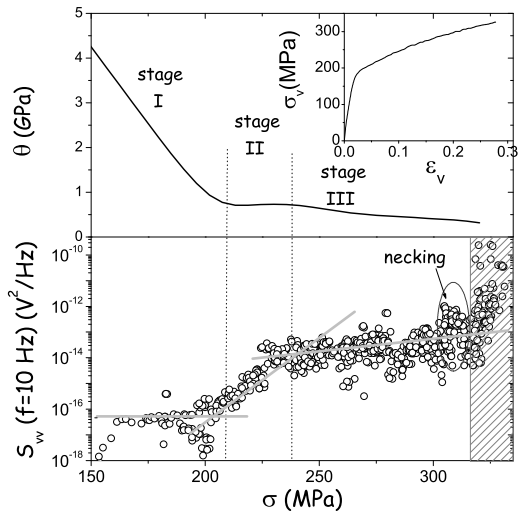}
\end{center}
\caption{(Color online)Top: Kocks-Mecking plot ( $\theta$  vs. $\sigma$) for Cu
 polycrystalline thin wire plastically strained in tension. Also shown are the typical stages of plastic deformation
 (The inset shows the corresponding stress-strain curve of the material).
Bottom: Voltage Noise at f=10 Hz as function of the strain.}
\label{fig.2}
\end{figure}

\begin{figure}[t!]
\begin{center}
\includegraphics*[width=8.0cm]{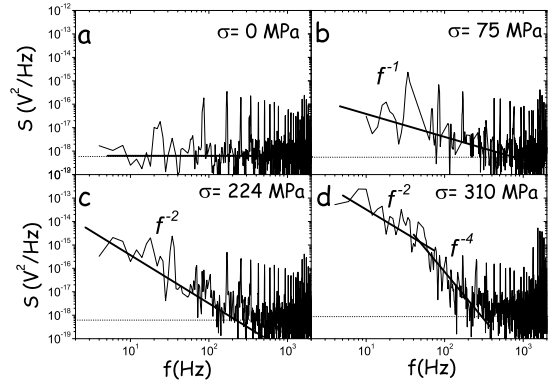}
\end{center}
\caption{(Color online) Typical spectra observed during the wire deformation, for different strain values.
 The appearance of a $f^{-4}$ dependence is consistent with dislocation clustering, observed here for strain values close to the necking.}
\label{fig.3}
\end{figure}

\begin{figure}[t!]
\begin{center}
\includegraphics*[width=8.0cm]{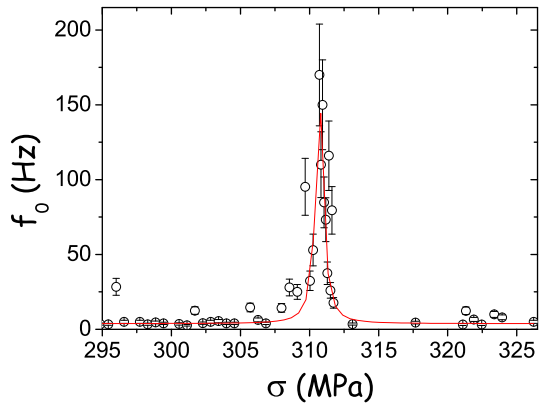}
\end{center}
\caption{(Color online) Variation of the cross-over frequency $f_0$ as function of the strain close to the necking.
 Note the maximum of $f_0$ at $\sigma \approx 311 MPa$.}
\label{fig.4}
\end{figure}

\begin{figure}[t!]
\begin{center}
\includegraphics*[width=8.0cm]{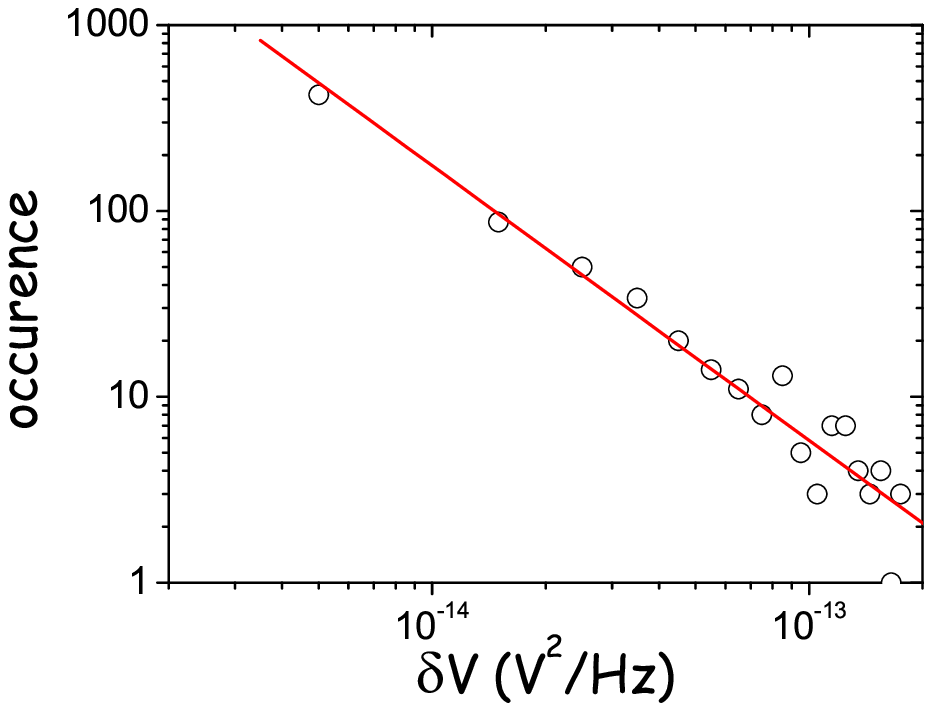}
\end{center}
\caption{(Color online) Histogram of the noise values in a log-log scale. The fit gives a slope $\alpha=$ -1.5 $\pm$ 0.1.}
\label{fig.5}
\end{figure}


\begin{references}
\label{sec:TeXbooks}
\bibitem{Miguel} M. C. Miguel, A. Vespignani, S. Zapperi, J. Weiss, and J. R. Grasso, Nature (London) 410, 667 (2001)
\bibitem{Zaiser} M. Zaiser, Adv. Phys. 55, 185 (2006).
\bibitem{Dimiduk} D. M. Dimiduk, C. Woodward, R. LeSar, and M. D. Uchic, Science 312, 1188 (2006).
\bibitem{Lebyo09} M. A. Lebyodkin, T. A. Lebedkina, F. Chmelík, T. T. Lamark, Y. Estrin, C. Fressengeas, and J. Weiss, Phys. Rev. B 79, 174114 (2009).
\bibitem{dislo} P.V. Andrews, M.B. West and C.R. Roheson, Phil. Mag. 19, 887 (1969).
\bibitem{Dislocationnoise} G. Bertotti, A. Ferro, F. Fiorillo and P. Mazetti in "Electrical noise associated with dislocations and plastic flow in metals",
Dislocations in Solids, Vol. 7 : F.R.N. Nabarro, ed. (North-Holland, Amsterdam), 1 (1986). 
\bibitem{DDH} P. Dutta, P. Dimon, and P. M. Horn, Phys. Rev. Lett. 43, 646 (1979).
\bibitem{Currentnoise} G. Bertotti, F. Fiorillo and P. Mazetti, Phys. Scripta T1, 134 (1982). 
\bibitem{Bert} G. Bertotti, M. Celasco, F. Fiorillo, and P. Mazzetti, J. Appl. Phys. 50, 6948 (1979).
\bibitem{Pat} Dinesh Patel, Yves Dumont, and I. L. Spain, J. Appl. Phys. 72, 1901 (1992).
\bibitem{Gior} D. M. Fleetwood and N. Giordano, Phys. Rev. B 28, 3626 (1983).
\bibitem{Haw} Lyndon D. Segales, James R. Gaines, Anupam K. Misra, and Richard E. Rocheleau, J. Appl. Phys. 88, 4146 (2000).
\bibitem{Narutani} T. Narutani and J. Takamura, Acta Metal. Mat. 39-8,2037 (1991).
\bibitem{jo} J. Scola, A. Pautrat, C. Goupil, and Ch. Simon, Phys. Rev. B 71, 104507 (2005)
\bibitem{Mecking 1977} H. Mecking, in: A.W. Thompson (Ed.), Work hardening in tension and fatigue, TMS-AIME, New-York, 67 (1977).
\bibitem{Flinn et al. 2001} J.E. Flinn, D.P. Field, G.E. Korth, T.M. Lillo, J. Macheret, Acta Mater. 49, 2065 (2001).
\bibitem{Essmann and Mughrabi 1979} U. Essmann, H. Mughrabi, Philos. Mag. A40, 731 (1979).
\bibitem{Sumino et al. 1963} K. Sumino, Y. Kawasaki, M. Yamamoto, M.P. Sumino, Acta Metall. 11, 1235 (1963).
\bibitem{Huang 1998} X. Huang, Scripta Mater. 38, 1697 (1998).
\bibitem{Keller et al. 2010} C. Keller, E. Hug, R. Retoux and X. Feaugas,, Mech. Mater.42, 44 (2010).
\bibitem{Mecking and Kocks 1981} H. Mecking and U.F. Kocks, Acta Metall.29, 1865 (1981).
\bibitem{Mughrabi 1987} H. Mughrabi, Mat.Sci.Eng. A 85,15 (1987).
\bibitem{vdz} A. Van der Ziel, Physica (Utrecht) 19, 742 (1953).
\bibitem{mazetti} G. Montalenti, Revue de Physique Appliqu\'ee 5, 87 (1970).
\bibitem{Weiss97} J. Weiss, T. Richeton, F. Louchet, F. Chmelik, P. Dobron, D. Entemeyer, M. Lebyodkin, T. Lebedkina,
 C. Fressengeas, and R. J. McDonald, Phys. Rev. B 76, 224110 (2007).           

\end{references}
\end{document}